\documentclass[journal,comsoc]{IEEEtran}
\usepackage{cite}
\usepackage[T1]{fontenc}% optional T1 font encoding
\usepackage{amsmath,amssymb}
\usepackage{graphicx}
\usepackage{textcomp}
\usepackage[dvipsnames]{xcolor}
\usepackage{epstopdf}
\usepackage{balance}
\usepackage[stable]{footmisc}

\usepackage[hyphens]{url}
\usepackage{hyperref}
\usepackage[hyphenbreaks]{breakurl}
\usepackage{enumitem}
\usepackage{makecell}

\DeclareMathOperator*{\argmin}{argmin}
\newcommand{\p}{p_{\bb{Y}|\textbf{X}}}

\newcommand{\bb}[1]{\mathbf{#1}}

\ifCLASSINFOpdf
\else
\fi
\usepackage{amsmath}
\usepackage[cmintegrals]{newtxmath}
\begin{document}
	
	\title{\vspace{-0.25cm}Impact of Bit Allocation Strategies on Machine Learning Performance in Rate Limited Systems}%Bit Allocation for Distributed Quantization with Central Intelligent Data Processing Unit}
	%Bit Allocation for Distributed Quantization and in-network Intelligent Functions
	%Bit Allocation for Quantization of Multiple Correlated Sources using Cross Entropy
	
	% 4 separate \thanks must be used for each paragraph as LaTeX2e's \thanks
	
	\author{\vspace{-0.3cm}Afsaneh~Gharouni,~\IEEEmembership{Member,~IEEE,}
		Peter~Rost,~\IEEEmembership{Member,~IEEE,}
		Andreas~Maeder,~\IEEEmembership{Member,~IEEE,}
		and~Hans~Schotten,~\IEEEmembership{Member,~IEEE.}% <-this % stops a space
		\vspace{-0.3cm}
		\thanks{Afsaneh~Gharouni, Peter~Rost and Andreas~Maeder are with Nokia Bell Labs, Munich, Germany.}
		\thanks{Hans~D.~Schotten is with Institute of Wireless Communication and Navigation, Technical University of Kaiserslautern, Kaiserslautern,  Germany.}
		\vspace{-0.6cm}}
	
	%of Electrical and Computer Engineering, Georgia Institute of Technology, Atlanta,
	

	\markboth{}%Journal of \LaTeX\ Class Files,~Vol.~, No.~, ~2020}%
	{Shell \MakeLowercase{\textit{et al.}}: Bare Demo of IEEEtran.cls for IEEE Communications Society Journals}




	\maketitle
	
	\begin{abstract}		
		Intelligent entities such as self-driving vehicles, with their data being processed by machine learning units (MLU), are developing into an intertwined part of networks. These units handle distorted input but their sensitivity to noisy observations varies for different input attributes. Since blind transport of massive data burdens the system, identifying and delivering relevant information to MLUs leads in improved system performance and efficient resource utilization. Here, we study the integer bit allocation problem for quantizing multiple correlated sources providing input of a MLU with a bandwidth constraint.  
		
		Unlike conventional distance measures between original and quantized input attributes, a new Kullback-Leibler divergence based distortion measure is defined to account for accuracy of MLU decisions. The proposed criterion is applicable to many practical cases with no prior knowledge on data statistics and independent of selected MLU instance. Here, we examine an inverted pendulum on a cart with a neural network controller assuming scalar quantization. Simulation results present a significant performance gain, particularly for regions with smaller available bandwidth. Furthermore, the pattern of successful rate allocations demonstrates higher relevancy of some features for the MLU and the need to quantize them with higher accuracy. 
		
	\end{abstract}
	\vspace{-.15cm}
	

	% \fi
	\IEEEpeerreviewmaketitle

	\vspace{-.45cm}
	\section{Introduction}
	\vspace{-.25cm}
	With increasing number of applications deploying connected devices to perform complicated tasks, machine learning based units (MLUs) become an integrated part of mobile networks. Hence, considering functionality of these blocks in design of communications systems is beneficial in order to both enhancing system performance and utilizing radio resources efficiently. 
	MLU input space contains attributes with different levels of relevance and redundancy regarding the output. Accordingly, severity of performance loss in response to corrupted inputs depends on relevancy of the features. Explaining this behavior is complicated, especially in presence of dependencies among input variables. To this end, we revisit the rate allocation problem and suggest an automated way to determine levels of distortion that MLU can tolerate while reducing its prediction errors given a bandwidth constraint. 
	
	The tradeoff between compression and accuracy is a well-known dilemma in lossy quantization. Due to the complexity of distributed scenarios, achievable rate distortion (RD) regions are derived for special cases. These studies can be categorized into syntax and relevance based solutions. The syntax based category presents approaches measuring the distance between source sequences $\mathbf{x}$ and their decoded versions $\hat{\mathbf{x}}$. The RD theory \cite{elementsof}, Wyner-Ziv coding and its network extension~\cite{WZ, Gastpar_WZ_ext}, quadratic Gaussian multiterminal source coding (MSC)~\cite{Complete_G_Multiterminal} and MSC for two encoders under logarithmic loss~\cite{The_Multiterminal} belong to this first group. 
	These solutions provide the basis for establishing reliable human to human communications. However, exact reconstruction of transmitted messages is not an optimal criterion when dealing with MLUs in network. In these cases, achieving a high accuracy on final outputs $\mathbf{y}$ determines the system performance.
	
	In order to consider final machine learning (ML) predictions in distortion measure, the second relevance based category of solutions target to compress $\mathbf{x}$ while preserving the relevant information for prediction of $\mathbf{y}$. These methods are also tailored for special cases assuming prior knowledge on statistical relation among random variables (RVs) or their probability distributions. Information bottleneck (IB) is a RD function compressing one RV $x$ in a single encoder-decoder system, where mutual information between the quantized message and another variable of interest $y$ is the  distortion measure~\cite{the_IB, IBRDproof}. The objective function of this optimization problem has also been used for quantization codebook design \cite{IB_codebook}.
	
	Several studies attempted to extend IB for distributed quantization with multiple sources. Multivariate IB introduced in \cite{MIB} employs Bayesian networks (BN) for this purpose, where the optimal assignment form is derived. However, the optimality of this proposal in terms of determining RD regions is not discussed, and its cost function has not been used to select number of clusters in literature. It should also be noted that BN determination is generally far from trivial for ML tasks. Authors of \cite{dist_dv_G} characterize the RD region of distributed IB for discrete and vector Gaussian sources assuming conditional independence of observations given the main signal of interest which does not hold in many learning problems.

	The Chief Executive Officer (CEO) problem studies the estimation of a data sequence using its independently corrupted versions observed by different agents~\cite{CEO}. These observations are quantized and communicated to a single decoder. The general formulation of CEO can be accounted as relevance based compression, however, its RD region is only investigated for special cases which are not applicable for learning paradigms. The Gaussian CEO \cite{QGCEO, VGCEO1, OptVGCEO} addresses corruptions caused by additive white Gaussian noise. This simple setup cannot comply with complicated MLU models. \cite{The_Multiterminal} provides the RD region of %two encoder MSC and 
	$m$-encoder CEO conveying information regarding another RV under logarithmic loss. Aside from the specific distortion measure having an important impact on making this problem tractable, as in all CEO setups conditional independence of observed sequences given the original data is assumed. 
	Considering the mentioned aspects, these CEO studies have not been evaluated for learning tasks. 	
	
	In addition, authors of \cite{localiz} study the problem of 1-bit rate allocation for localization in wireless sensor networks, while the proposed cost function accounts for both decoding and localization error. 
	
	Fixed-rate quantization has three main aspects: rate allocation, codebook design, and assignment of RVs to codewords.  
	Here, we focus on integer-valued bit allocation for multiple correlated sources performing scalar uniform quantization with arbitrary distributions while MLU is treated as a black box. This includes all non-adaptive ML blocks once trained and executing tasks online in network, independent of their hypothesis and learning paradigm such as supervised and reinforcement learning,  e.g., the proposed approach can be applied on \cite{fr1, fr2} after the convergence. Thus, the provided solution can be used in a wide variety of real-world scenarios. 
	
	In this paper, we propose a criterion using Kullback-Leibler divergence (KLD) to measure quality of bit allocations. 
	The KLD approximation is performed and discussed for two non-parametric approaches: histogram with smoothing and k-nearest neighbor (kNN). 
	Then performance of the proposed method is investigated for a cart inverted pendulum with ML based controller (MLC) which is a shallow neural network (NN). The results are compared with those of equal bit sharing and a mean square error (MSE) based approach inspired by asymptotically optimal integer-valued bit allocation for Gaussian distributed RVs from~\cite{Integer_opt}. Simulation results show significant gain in system performance for low bit rate region. It can also be seen that a lower quantization noise can be tolerated on two of the features compared to other RVs. \footnote{The proposed method has also shown significant gains for other use-cases including a different setup for the inverted pendulum, indoor environment classification with real data and a synthetic data set. These simulations are presented in an extended version of the paper on arXiv.}

	This paper is organized as follows. The system model is discussed in Section~\ref{SysModel}. In Section~\ref{method}, the rate allocation approach and KLD estimators are introduced. The simulation setup is elaborated in Section~\ref{EvalSetup}, and numerical results are presented\hspace{-.1cm} in Section\hspace{-.1cm} \ref{numeric}. Finally, \hspace{-.05cm}conclusions are drawn in  Section~\ref{conclusion}.

	\textit{Notation}: Linear-quadratic regulator (LQR) controller matrices $\bb{K}, \bb{Q}$ and vectors are typeset boldface. $\mathbf{x} = [x_1, \cdots, x_N]$ and $\hat{\mathbf{x}} = [\hat{x}_1, \cdots, \hat{x}_N]$ are vectors of  non-quantized and quantized MLU input components, and $\mathbf{y}$ represents MLU output. The $i$th element of these vectors is denoted with subscript $i$ as in $x_i$. %The input vector of linear-quadratic regulator (LQR) controller with fixed bar mass and length is shown as $\bb{x}_{LQR}$. For the inverted pendulum problem that we considered, $$. 
	$p_{\hat{\bb{X}}, \bb{Y}}(\hat{\bb{x}}, \bb{y})$ also shown as $p$, stands for the joint input-output distribution of the MLU assuming a highly accurate quantization. The joint MLU input-output distribution for a given bit allocation $\bb{R} = \{ R_i \}$ is shown as $q_{\hat{\bb{X}}, \bb{Y}}(\hat{\bb{x}}, \bb{y})$ or simply $q$.  Data set samples for estimation of KLD are indicated as $\bb{z}_j = [\hat{\bb{x}}_j, \bb{y}_j ] $. Finally, $\hat{p}(\bb{z}_j)$ and $\hat{q}(\bb{z}_j)$ are distribution estimations for $p$, $q$ with data set samples.

	%\IEEEPARstart{T}{his} demo file is intended to serve as a ``starter file''
	%for IEEE Communications Society journal papers produced under \LaTeX\ using IEEEtran.cls version 1.8b and later.
	%I wish you the best of success.
	
	

	\vspace{-.4cm}	
	\section{System Model} \label{SysModel}
	\vspace{-.15cm}
	\subsection{General Description}
\begin{figure}[t!]
		\centering
		\includegraphics[width=.85\columnwidth]{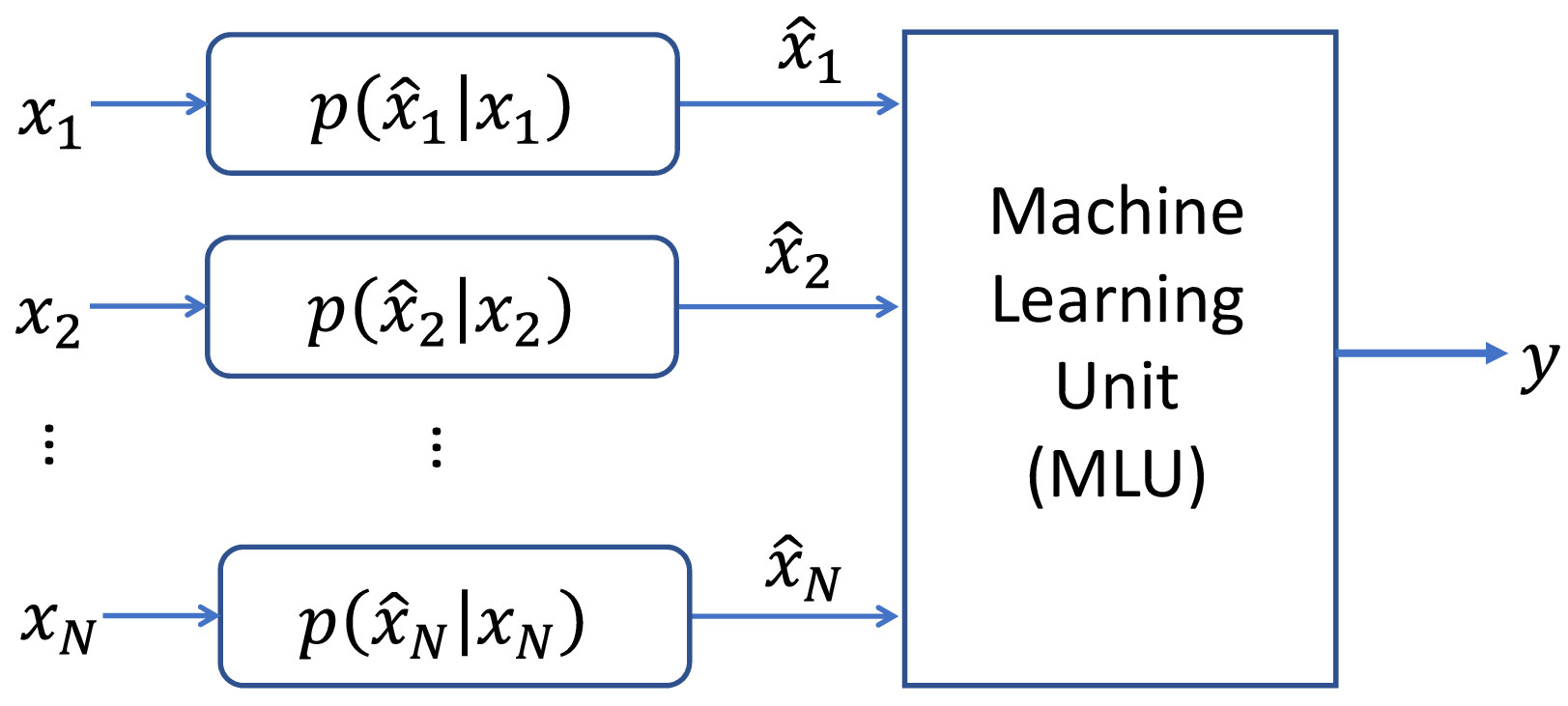}
		\vspace{-.3cm}
		\caption{Block diagram of the system model.}
		\label{BlockDiag}
		\vspace{-.5cm}
	\end{figure}

	\vspace{-.1cm}
	As shown in Fig.~\ref{BlockDiag}, we study a multiple access channel scenario in which $N$ memoryless stationary sources provide real-valued input attributes $\mathbf{x}$ for a MLU. In presence of complex-valued attributes, the real and imaginary parts can be separated and treated as different RVs. The system performance is evaluated in terms of accuracy on predicting MLU output values $\bb{y}$. The scalar uniform quantization with $R_i$ bits for each symbol is performed on RV of $i$th source which is shown as $p_{\hat{X_i}|X_i}(\hat{x_i}|x_i)$. It is assumed that quantized vector is received error-free at the receiver. Note that application of the proposed method is not dependent on this assumption. To remove it, $\hat{\mathbf{x}}$ should be redefined to capture the effect of factors such as channel coefficient and  receiver noise. 
	Here, we seek to build a system model that can be used in practice. So, with no further assumptions, input attributes can be highly correlated and have an arbitrary joint probability density function $p_{\mathbf{X}}(\mathbf{x})$ with $\bb{x} \in \mathcal{X}_1 \times \mathcal{X}_2 \times \cdots  \mathcal{X}_N $. 
	
	Given the available bandwidth $B$ and signal to noise ratio (SNR) $\gamma=\frac{E_b}{N_0 B T_b}$, 
	where $E_b$, $T_b$ and $N_0$ are energy per bit, bit interval and noise power spectral density, respectively, the capacity of bandlimited channel is $C_B = B \times \log_2(1+\gamma)$  bits/sec. Thus, the constraint for allocating bandwidth $B_i$ to $i$th source is $\sum_i B_i \leq B$. Assuming same SNR for all terminals, $\gamma_i = \gamma$, and a given symbol interval $T_s$, the constraint becomes $\sum_i R_i \leq R_\mathrm{sum}$, where $R_i = B_i \times \log_2(1+\gamma) \times T_s$ is the number of bits quantizing each symbol of the $i$th terminal, and $R_\mathrm{sum} = C_B \times T_s$~bits for each symbol~interval. $R_i$ is assumed to be integer-valued as usual in practical systems. The set of feasible bit allocations meeting the constraint are shown by $\mathcal{R}$. To consider different SNR values, the corresponding possible bit allocations should be added to the feasible set.

	In many scenarios, training is performed independent of communications system design and we are not able to modify the MLU. Therefore, it is assumed that learning process is done by non-quantized data and MLU parameters are fixed. In this case, $\sum_i R_i >> R_\mathrm{sum}$ and the joint probability distribution on input and output of the MLU is $p_{\hat{\bb{X}}, \bb{Y}}(\hat{\bb{x}}, \bb{y})$ which is also stated as $p_{\bb{X}, \bb{Y}}(\bb{x}, \bb{y})$ to simplify the notation. This distribution is considered as the true distribution and is used as reference to perform comparisons. 
	
	Since the MLU model is trained and fixed, and following Markov chain of the system $\bb{Y} \leftrightarrow \bb{X} \leftrightarrow \hat{\bb{X}}$, we can write $q_{ \bb{Y}|\hat{\bb{X}}}(\bb{y} |\hat{\bb{x}}) = \sum_{\bb{x}^\prime \in \mathcal{X}^N} \p(\bb{y}|\bb{x}^\prime) p_{\bb{x} | \hat{\bb{x}}}(\bb{x}^\prime | \hat{\bb{x}})$ or equivalently, \mbox{$q_{ \bb{Y}|\hat{\bb{X}}}(\bb{y} |\hat{\bb{x}}) = \frac{1}{q_{\hat{\bb{X}}}(\hat{\bb{x}})} \sum_{\bb{x}^\prime \in \mathcal{X}^N} p_\bb{X}(\bb{x}^\prime) p_{\hat{\bb{X}}|\bb{X}}(\hat{\bb{x}}|\bb{x}^\prime) \p(\bb{y}|\bb{x}^\prime)$, where} $\p(\bb{y}|\bb{x}^\prime)$ is the fixed distribution learned by ML, and distribution of quantized data $q_{\hat{\bb{X}}}(\hat{\bb{x}})$ and conditional distributions on $\bb{x}$ and $\hat{\bb{x}}$ change for different rate allocations. 
	
	In order to compare our results with syntax based solutions, a typical MSE based approach is considered. The selected bit allocation using this method is 
	\vspace{-.4cm}
	\begin{equation} \label{mse}
	\mathbf{R}^* = \argmin_{\bb{R} \in \mathcal{R} } \sum_{i=1}^N \sigma_i^2, 
	\vspace{-.15cm}
	\end{equation} 
	where $\sigma_i^2 = \mathbb{E}_{x_i} \{ (x_i- \hat{x_i})^2 \}$ is the MSE between $i$th input feature and its quantized version which is calculated by employing data sets. Expectation is denoted by $\mathbb{E} \{ \cdot \}$.
	
	Equal sharing is another method that we investigate to provide a comparison baseline. In this case, $R_i = \lfloor R_\mathrm{sum} / N \rfloor$ and $\lfloor \cdot \rfloor$ returns the greatest integer which is equal or less than its input. This choice of $R_i$ complies with our assumption on no exchange of knowledge among sources and integer-valued $R_i$. Hence, $R_i$ changes only if remainder of $R_\mathrm{sum}/N$ is zero.

	\vspace{-.5cm}
	\subsection{Inverted Pendulum on Cart}
	\vspace{-.2cm}
	In order to evaluate performance of bit allocations, we investigate the  control problem of inverted pendulum on a cart. The controller is supposed to move the cart to position $r = 0.2$ meter in less than 2 seconds while the pendulum is in its equilibrium position, i.e., $\theta= 0$, where $\theta$ is the angle of pendulum with respect to vertical axis. The initial deviation from vertical position is between $-0.1$ and 0.1 radians while the pendulum is placed at $r = 0$. For a given bar length and mass, steady state equations governing the plant are given by
	\vspace{-.2cm}
	\begin{equation} \label{LQR}
	\dot{\bb{x}}_\mathrm{LQR}^\mathrm{T}
	= \begin{bmatrix} 0 & 0 & 1 & 0 \\ 0 & 0 & 0  & 1\\  0 & \frac{m^2 g l^2 }{c} & \frac{-b(I+ml^2)}{c} & 0\\ 0 & \frac{mgl(M+m)}{c}  & \frac{-mlb}{c} & 0\\ \end{bmatrix} 
	\bb{x}_\mathrm{LQR}^\mathrm{T} + 
	\begin{bmatrix} 0 \\ 0 \\\frac{I+ml^2}{c} \\ \frac{ml}{c} \end{bmatrix} f,
	\vspace{-.15cm}
	\end{equation}
	where $\bb{x}_\mathrm{LQR} = [r, \theta, \dot{r}, \dot{\theta}] $, $\dot{\bb{x}}_\mathrm{LQR}$ is its derivative with respect to time. $c=(M+m)I+mMl^2$ with $M$, $m$ and $l$ being the cart mass, pendulum mass and length to pendulum center of mass, respectively. $I=ml^2/3$ stands for the moment of inertia for bar mass. $g = 9.8$ and $b = 0.1$ (N/m/sec) are assumed as standard gravity and coefficient of friction for the cart. Finally, $f$ is the force applied to the cart in horizontal direction. 
	
	To calculate the optimal force, LQR controller with precompensation factor is used for different values of bar length and mass. The cost function of LQR is $\int \bb{x}_\mathrm{LQR}^\mathrm{T} \bb{Q} \bb{x}_\mathrm{LQR} + \bb{u}^\mathrm{T} R_\mathrm{LQR} \bb{u}$, where $\bb{u} = - \bb{K} \bb{x}_\mathrm{LQR}$ and $\bb{K}$ is the matrix of controller coefficients. $\bb{Q}$ and $R_\mathrm{LQR}$ are controller parameters to balance the relative importance of error and control effort, e.g., energy consumption.
	
	The system performance of this problem is evaluated in terms of steady state errors. The error-bands for cart position and angle of pendulum are $0.1$ meters and $0.01$ radians, respectively. Thus, an error is counted when the deviation from equilibrium position is outside of these intervals in the last 100 milliseconds, e.g., $|\theta_\mathrm{final}| > 0.01$. Considering steady state error is a standard way of evaluating controllers in a predefined period of time. A steady state error can occur while the system becomes stable after the aforementioned 2 seconds.

	\vspace{-.4cm}
	\section{Kullback-Leibler Divergence Based Bit Allocation and its Estimation} \label{method}
	\vspace{-.15cm}
	The goal is to find the bit allocation set $\mathbf{R}^* $ which minimizes the following cost function	
	\vspace{-.2cm}
	\begin{equation}
	\label{DKL}
	\bb{R}^* = \argmin_{ \bb{R} \in \mathcal{R} } D_{\mathrm{KL}} \Big(   p_{\hat{\bb{X}}, \bb{Y}}(\hat{\bb{x}}, \bb{y}) ||q_{\hat{\bb{X}}, \bb{Y}}(\hat{\bb{x}}, \bb{y}) \Big),
	\vspace{-.1cm}
	\end{equation}
	where $D_{\mathrm{KL}}(\cdot||\cdot)$ is the Kullback-Leibler divergence or relative entropy measuring dissimilarity between two distributions.
	$\mathcal{R}$ contains all the rate allocations satisfying $\sum_{i=1}^N R_i \leq R_\mathrm{sum}$, where $R_i > 0$ is an integer-valued number. To solve this optimization problem, we estimate the two distributions empirically as explained in the following.

	The quality and accuracy of solution provided by (\ref{DKL}) is highly dependent on KLD approximation accuracy. Here, we estimate $p$ and $q$ using non-parametric methods, histogram with smoothing and kNN. The histogram estimator is a simple approach with the drawback of having many bins with zero samples. In addition, number of its required bins increases exponentially with data dimension. We also consider kNN estimator to investigate the effect of DKL approximation accuracy on system performance. 
	kNN has been used for mixed continuous-discrete setups, and a high accuracy for strongly correlated data is not guaranteed for this estimator~\cite{kNN_dep}. Let each $T_1$ and $T_2$ be data sets containing $J$ samples $\{\bb{z}_j; j = 1, \cdots, J \}$ drawn from distributions $p$ and $q$, respectively. The kNN estimation of $p$ is
	\vspace{-.15cm}
	\begin{equation}
	\hat{p}(\bb{z}_j) = \frac{k}{J} \times \frac{1}{v(\bb{z}_j)}; \hspace{0.1cm} \bb{z}_j \in T_1,
	\vspace{-.2cm}
	\end{equation}
	where $v(\bb{z}_j) = \frac{\pi^{d/2}}{\Gamma(d/2 +1) } \times \frac{1}{R_p(\bb{z}_j)^d}$ is the volume of a $d$-dimensional ball with radius $R_p(\bb{z}_j)$. $\Gamma (\cdot)$ is the gamma function and $R_p(\bb{z}_j)$ stands for the euclidean distance between $\bb{z}_j$ and its $k$th neighbor in $T_1$. The $k$th neighbor of $\bb{z}_j$ is the $k$th sample in the list of sorted samples of $T_1$ from minimum to maximum euclidean distance regarding $\bb{z}_j$. $d$ is the sum of $N$ and dimension of ML outputs $\bb{y}$. Similarly, an estimate of $q$ can be calculated, where $R_q(\bb{z}_j)$ is the euclidean distance between $\bb{z}_j \in T_1$ and its $k$th neighbor in $T_2$. Therefore, the plugin estimator for KLD of (\ref{DKL}) becomes	
	\vspace{-.15cm}
	\begin{equation} \label{5}
	D_{\mathrm{KL}} ( p || q ) \approx \mathbb{E}_{\bb{z}}
	\Big\{ \log \Big(\frac{\hat{p}(\bb{z}_j) }{\hat{q}(\bb{z}_j)  } \Big) \Big\}.
	\vspace{-.15cm}
	\end{equation}

	A well-known difficulty with computing KLD is that to get a finite value, the support set of true distribution should be contained in support set of estimated distribution. While this is reasonable in some applications, it is an extreme condition for learning problems, particularly since distributions are only approximated with limited number of samples. Therefore, data smoothing can be used to overcome the problem. To deal with this situation, the width of histogram bins are selected to be larger than that provided by quantization. Thus for each sample in support set of $p$, we assume the existence of at least one sample when approximating $q$. In this case, instead of $\hat{q}(\bb{z}_j) = \frac{n}{J}$, where $n$ is the number of samples in histogram bin of $\bb{z}_j$, we have
	\vspace{-.15cm}
	\begin{equation}
	\hat{q}(\bb{z}_j) = \frac{n + \alpha}{J + \mu},
	\vspace{-.15cm}
	\end{equation} 
	where $\mu$ is the number of bins in support of $p$ with zero samples from $T_2$. For $n=0$, $\alpha=1$ and otherwise, $\alpha=0$. It is worth mentioning that in this rate allocation setup, the relative KLD values and their order are decisive, not absolute values. 
	
	The feasible set of this problem is non-convex due to the integer-valued bit allocation assumption, however, it contains a limited number of members. Thus for focusing on impact of KLD approach and its approximation on MLU output, estimations of (\ref{5}) are substituted in (\ref{DKL}) for members of $\mathcal{R}$ and a brute-force search finds the optimal solution. 
	
	In a high dimensional space, large number of required samples for meaningful estimations with a simple histogram can be restrictive. kNN method can circumvent this problem. The required kNN computations are theoretically expensive for a large data set. However, the calculations for both KLD approximations and solving (3) are performed only once and offline. Once the bit allocations are determined for different bandwidth constraints, one of them is picked for quantization according to the available bandwidth. Therefore, dealing with these computations is feasible in practice without affecting applicability of the proposed approach.

	%\vspace{-.35cm}
	\section{Simulation Setup} \label{EvalSetup}
	\vspace{-.05cm}
	\subsection{Training the MLC}
	\vspace{-.05cm}
	As the MLC, we train a fully-connected shallow NN with 70 neurons. The input features for MLC are mass and length of the bar pendulum, position $r$, velocity $v = \dot{r}$, angular position $\theta$ and angular velocity $q = \dot{\theta}$, implying an input layer dimension of 6. Hence, $\mathbf{x} = [m, l, r, \theta, v, q]$, where values of $m$ and $l$ can be selected from the ranges 0.1 to 2 kg and 20 to 50 cm, respectively. In addition, the output of MLC $y$ is the horizontal force applied to the cart which is shown as $f$ in (2).  The NN is trained with a data set generated using LQR controllers for different random values of bar mass and length, with the following parameters: $M=0.5$ kg, $R_\mathrm{LQR} = 0.1$ and $\bb{Q}$ is a $4 \times 4$ matrix with zero entries except for the first and third diagonal elements being 5000 and 100, respectively. The LQR parameters are selected based on a trial and error procedure as elaborated in~\cite{LQR}.
	The sampling time is $0.01$ seconds. The training and test set contain 600 and 200 sequences, each of length 200, respectively. Validation ratio is $\frac{1}{3}$\footnote{The proposed approach attempts to preserve the performance level of the given fixed MLU. Thus, common learning challenges such as having a limited number of training samples can only worsen the induced MLU performance which persists even in case of delivering nonquantized data. But such degradation is caused by the MLU itself and not the selected quantization.}.

	Here, we deal with a regression problem. Sigmoid and linear activation functions are used in hidden and output layer, respectively. MSE is the loss function for training and NN weights are initialized with Xavier uniform initializer. Batch gradient descent with batch size of 1000 is the search algorithm. Furthermore, the learning rate is 0.01 with no decay factor. Stop condition is getting no improvement in validation loss for 50 epochs which occurred after 641 epochs. The final MSE achieved on the test set is $\approx 0.23$.

	\vspace{-.45cm}
	\subsection{KLD Estimation and Rate Allocation}	
	\vspace{-.05cm}	
	We use the MLC to generate data sets for estimation of KLD.  For the uniform quantization, minimum and maximum values of each RV is taken from $T_1$. Since $m$ and $l$ are not expected to change frequently, we assume that their values are transmitted with 10 bits for each feature when needed. Members of $\mathcal{R}$ are selected to satisfy $3 \leq R_i \leq 9$ and $\sum_i R_i = R_\mathrm{sum}$, where we have $ R_\mathrm{sum} - 20$ bits to quantize the last four attributes of vector $\bb{x}$ described in \ref{EvalSetup}. This interval choice both limits the search space and is sufficiently large considering the range of RVs in this problem. For estimating $p$ and $q$, 40000 samples and the typical value of $k= \sqrt{J} = 200$ are used.

	As explained in section~\ref{SysModel}, we assume $ p_\bb{X}(\bb{x})$ is fixed which is the case for many non-adaptive learning problems. Thus, data set $T_2$ can be constructed directly from $T_1$ by simply quantizing its input samples for a given rate allocation and feeding them into the MLU to compute corresponding outputs. This procedure reduces computations significantly, because the alternative is to run simulations for pendulum environment to build a data set for each bit allocation. 
	
	On the other hand, for the specific problem of inverted pendulum, very low quality quantization results in force decisions with large distance from the true ones. And after feeding back these force decisions to the plant, $p_\bb{X}(\bb{x})$ starts to diverge from the assumed distribution and consequently, $T_1$ must be updated. In order to avoid this difficulty, distribution on $\bb{x}$ is estimated for different sum rate constraints and bit allocations. Then, KLD between distribution of these allocations and the true distribution $p_\bb{X}(\bb{x})$ is calculated. These KLD estimations show a small value for $R_\mathrm{sum} \geq 42$. Therefore, it is a valid assumption that $p_\bb{X}(\bb{x})$ is almost fixed for sum rate constraints larger than 42 bits.

	\vspace{-.55cm}
	\section{Numerical Result} \label{numeric}
	\vspace{-.1cm}
	\begin{figure}[t!]
		\centering
		\includegraphics[width=.95\columnwidth]{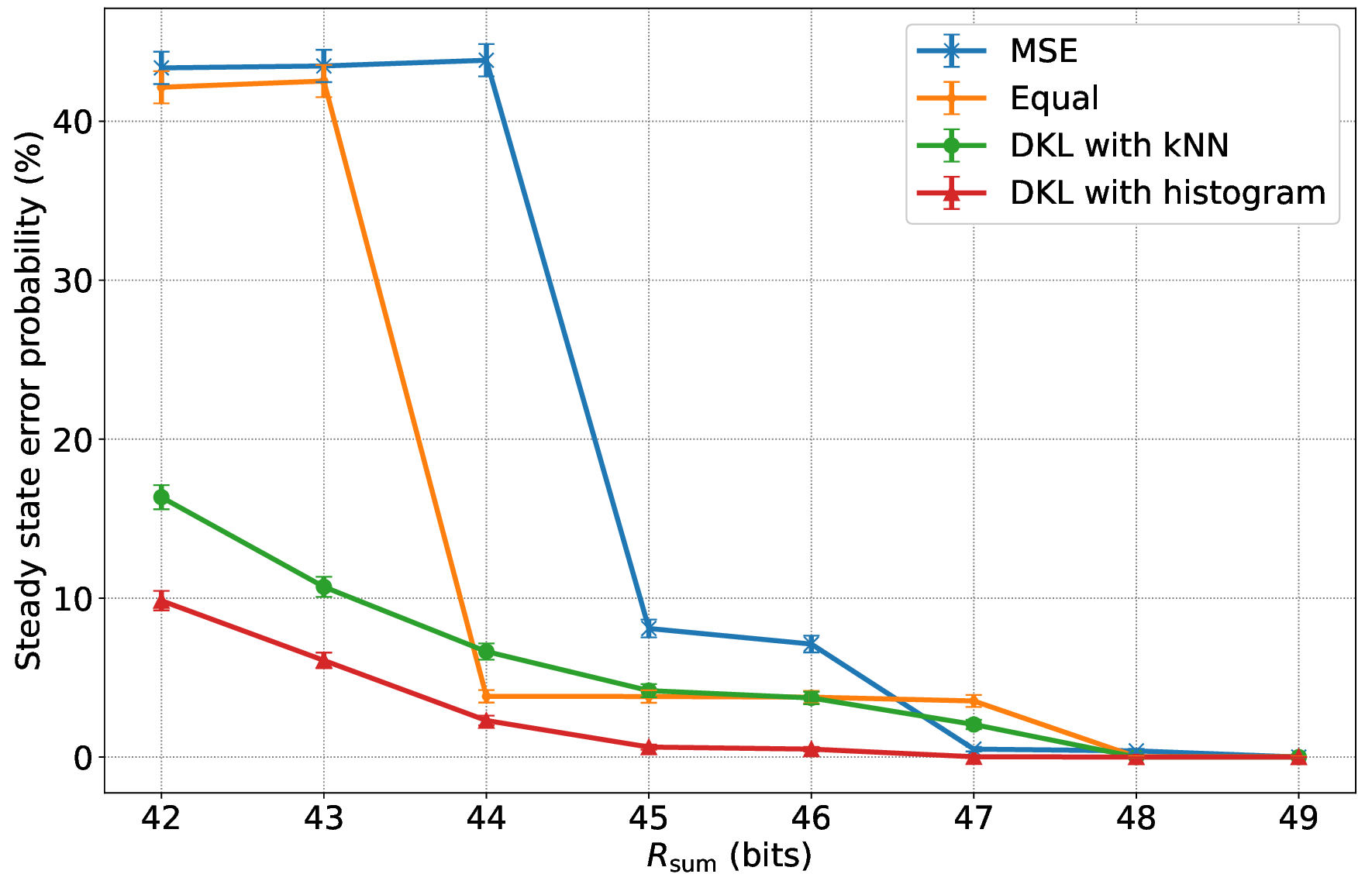}
		\vspace{-.35cm}
		\caption{Steady state error probability in percentage vs. $R_\mathrm{sum}$ the total number of quantization bits used in a symbol interval.}
		\label{SS}
		\vspace{-.55cm}
	\end{figure}

	In this section, the step response of cart inverted pendulum is monitored for 10000 iterations while each iteration simulates a period of 2 seconds. The steady state error probability $P_\mathrm{e}$ with confidence intervals derived by Wald method vs. total number of quantization bits used in a symbol interval $R_\mathrm{sum}$ is depicted in Fig.~\ref{SS}. Simulations are performed for the proposed KLD based approach with histogram and kNN estimation, equal bit sharing and MSE based rate allocation of (\ref{mse}).
	The proposed method with histogram estimation outperforms other techniques for all sum rate constraints, and indicates a gain of 2 bits in achieving $P_\mathrm{e}<0.001$ at 47 bits with respect to equal sharing and MSE methods. It should be noted that this single inverted pendulum scenario is a sandbox, and the gains and rate of the communication scheme in a real environment with signal overheads and more devices increases rapidly. Particularly, the KLD with histogram picks a significantly better bit allocation for low sum rate values. For instance, if 42 bits can be assigned for the system, error probability for both eqaul sharing and MSE are larger than $40\%$. This number can be reduced to $\approx10\%$ implying a reduction of more than $30\%$ in failures using the KLD. This huge gain is a result of taking ML output into consideration. 
	
	In order to study the distribution of quantization noise and its pattern when a low error probability is achieved, consider the KLD approach with histogram at 46 bits and $P_\mathrm{e}\approx0.005$.
	With this constraint, the number of allocated bits for features of $\mathbf{x}$ are $[10, 10, 6, 6, 6, 8]$. Assuming that quantization error variance is defined as $\sigma^2_i = \mathbb{E} \{ (x_i - \hat{x}_i)^2 \}$ for each feature, we have $\sigma^2_3 \approx \sigma^2_4 $ of order of $10^{-6}$. 
	However for $v$ and $q$, quantization variances are $\sigma^2_5 \approx 0.0003 $ and $\sigma^2_6 \approx 0.0001$ which are almost 100 times larger than that of $r$ and $\theta$. This pattern of having lower quantization noise for $\theta$ and $r$ remains the same for bit allocations which turn out to provide low probabilities of error. Therefore, it can be concluded that these features have a higher relevancy or importance for the MLU. 
	
	For $R_\mathrm{sum} \leq 46$, rate allocations selected by MSE criterion result in the worst steady state error performance among all the methods under study. This performance gap is larger at lower sum rate values, e.g., a loss of $37.4\%$ and $32.7\%$ at $R_\mathrm{sum}= 43$ regarding the KLD with histogram and kNN, respectively. Furthermore, MSE based technique shows a huge improvement from $R_\mathrm{sum} = 44 $ to 45 bits. The reason lies behind the range from which input features take their values, and the fact that MSE is calculated independent of MLC output. In this setup, $v$ and $q$ values are picked from intervals which are almost $9$ and $21$ times bigger than those of $\theta$ and $r$. Therefore at the beginning, the syntax based MSE allocates more bits for $q$ and $v$, although high accuracy on these less relevant RVs doesn't improve the force decision. The first significant enhancement only occurs when $\sigma^2_5$ and  $\sigma^2_6$ are small enough, so, extra bits are used for $\theta$. Thus, a change from 4 to 5 in number of bits for $\theta$ when $R_\mathrm{sum} = 44$ becomes 45 bits leads to a decrease of $\approx35.7\%$ in probability of error.  
	The second decrease is also a consequence of allocating 5 bits instead of 4 bits for $r$ when moving from $R_\mathrm{sum} = 46$ to 47.
	
	Equal sharing outperforms the MSE results given that $R_\mathrm{sum} \leq 46$, e.g., $P_\mathrm{e}$ $\approx 42.5\%$ instead of $\approx 43.5\%$ for 43 bits. As stated before, the rate allocation provided by this method remains the same, unless sum rate is divisible by 4 which explains improvements at 44 and 48 bits. This method provides better results than KLD with kNN for the constraint of 44 which can be interpreted as a lucky situation for this approach. With 44 bits, equal sharing allocates 6 bits for each of $r, \theta, v$ and $q$. This indicates less quantization noise for more relevant RVs $\theta$ and $r$ which only happens because of their smaller intervals in this specific pendulum scenario. On the other hand, KLD with kNN is not capable of following distributions accurately and settles for a worse bit allocation with $\approx  3\%$ more failures than that of equal sharing.
	
	As expected, changing histogram estimator to kNN degrades the performance since kNN is not capable of providing a highly accurate estimation of KLD, particularly for the system under investigation with highly correlated variables. 
	However, it still offers less number of errors compared with the MSE approach for $R_\mathrm{sum} \leq 46$. For the constraint with 42 bits, it achieves a gain of $27\%$ and $25.8\%$ in comparison to MSE and equal bit sharing methods but the selected rate allocation causes $\approx 6.5\%$ higher error probability with respect to the KLD with histogram estimator. KLD with kNN also provides a better or equivalent performance regarding equal sharing for most points, except for $R_\mathrm{sum} = 44$ which was discussed.

	As shown by the numerical results, using the relevance based KLD approach with histogram is more beneficial in terms of fulfilling the requirements imposed by ML functionalities in a bandwidth limited system. In operation points with high probability of stability, the quantization noise on angle and position are much smaller than other features which indicates they have a higher level of relevance for the MLU. This knowledge can be used in case of having limited resources for providing a best-effort performance.

	\vspace{-.45cm}
	\section{Conclusion} \label{conclusion}
	\vspace{-.1cm}
	Since intelligent elements governed by ML become an integrated part of communications networks, we introduced a KLD based rate allocation for quantization of multiple correlated sources delivering input of a MLU. Simulation results show that the proposed method provides promising gains in system performance of a cart inverted pendulum problem, particularly for more restricted bandwidth constraints. It should be noted that the final outcome is use-case dependent and more importantly, it highly relies on KLD estimation accuracy. These observations motivate the shift from syntax to relevance based designs which operate in accordance with MLU requirements considering rate and resource limitations. Some potential problems to be addressed in future are to introduce low complexity methods for dealing with instantaneous fluctuations in channel quality, and studying of iterative algorithms to improve the overall system performance by targeting the combination of codebook design, assignment and bit allocation.

	\vspace{-.45cm}
	\bibliographystyle{IEEEtran}  
	\bibliography{DKL_Rate_alloc_paper_refs, Review_major_bib} 

\begin{thebibliography}{10}
\providecommand{\url}[1]{#1}
\csname url@samestyle\endcsname
\providecommand{\newblock}{\relax}
\providecommand{\bibinfo}[2]{#2}
\providecommand{\BIBentrySTDinterwordspacing}{\spaceskip=0pt\relax}
\providecommand{\BIBentryALTinterwordstretchfactor}{4}
\providecommand{\BIBentryALTinterwordspacing}{\spaceskip=\fontdimen2\font plus
\BIBentryALTinterwordstretchfactor\fontdimen3\font minus
  \fontdimen4\font\relax}
\providecommand{\BIBforeignlanguage}[2]{{%
\expandafter\ifx\csname l@#1\endcsname\relax
\typeout{** WARNING: IEEEtran.bst: No hyphenation pattern has been}%
\typeout{** loaded for the language `#1'. Using the pattern for}%
\typeout{** the default language instead.}%
\else
\language=\csname l@#1\endcsname
\fi
#2}}
\providecommand{\BIBdecl}{\relax}
\BIBdecl

\bibitem{elementsof}
T.~M. Cover and J.~A. Thomas, \emph{Elements of Information Theory (Wiley
  Series in Telecommunications and Signal Processing)}.\hskip 1em plus 0.5em
  minus 0.4em\relax USA: Wiley-Interscience, 2006.

\bibitem{WZ}
A.~{Wyner} and J.~{Ziv}, ``The rate-distortion function for source coding with
  side information at the decoder,'' \emph{IEEE Trans. on Information Theory},
  vol.~22, no.~1, pp. 1--10, January 1976.

\bibitem{Gastpar_WZ_ext}
M.~{Gastpar}, ``On {Wyner-Ziv} networks,'' in \emph{The Thrity-Seventh Asilomar
  Conf. on Signals, Systems Computers}, vol.~1, Nov 2003, pp. 855--859.

\bibitem{Complete_G_Multiterminal}
A.~B. {Wagner}, S.~{Tavildar}, and P.~{Viswanath}, ``Rate region of the
  quadratic gaussian two-encoder source-coding problem,'' \emph{IEEE Trans. on
  Information Theory}, vol.~54, no.~5, pp. 1938--1961, 2008.

\bibitem{The_Multiterminal}
T.~A. {Courtade} and T.~{Weissman}, ``Multiterminal source coding under
  logarithmic loss,'' \emph{IEEE Trans. on IT}, vol.~60, no.~1, 2014.

\bibitem{the_IB}
N.~Tishby, F.~C. Pereira, and W.~Bialek, ``The information bottleneck method,''
  \emph{Proc. 37th Annual Allerton Conf. on Comm., Control, and Computing}, p.
  368–377, 1999.

\bibitem{IBRDproof}
R.~Gilad-bachrach, A.~Navot, and N.~Tishby, ``An information theoretic tradeoff
  between complexity and accuracy,'' in \emph{In Proceedings of the
  COLT}.\hskip 1em plus 0.5em minus 0.4em\relax Springer, 2003, pp. 595--609.

\bibitem{IB_codebook}
S.~{Lazebnik} and M.~{Raginsky}, ``Supervised learning of quantizer codebooks
  by information loss minimization,'' \emph{IEEE Trans. on Pattern Analysis and
  Machine Intelligence}, vol.~31, no.~7, pp. 1294--1309, 2009.

\bibitem{MIB}
N.~{Slonim}, N.~Friedman, and N.~Tishby, ``Multivariate information
  bottleneck,'' \emph{Neural computation}, vol.~18, pp. 1739--89, Sep. 2006.

\bibitem{dist_dv_G}
\BIBentryALTinterwordspacing
I.~E. Aguerri and A.~Zaidi, ``Distributed information bottleneck method for
  discrete and gaussian sources,'' \emph{International Zurich Seminar on
  Information and Communication}, pp. 35--39, Feb. 2018. [Online]. Available:
  \url{https://doi.org/10.3929/ethz-b-000245048}
\BIBentrySTDinterwordspacing

\bibitem{CEO}
T.~Berger, Z.~Zhang, and H.~Viswanathan, ``The {CEO} problem [multiterminal
  source coding],'' \emph{IEEE Trans. on IT}, vol.~42, no.~3, 1996.

\bibitem{QGCEO}
H.~{Viswanathan} and T.~{Berger}, ``The quadratic gaussian {CEO} problem,'' in
  \emph{Proceedings of IEEE International Symposium on IT}, 1995, p. 260.

\bibitem{VGCEO1}
Y.~{Ugur}, I.~E. {Aguerri}, and A.~{Zaidi}, ``Vector gaussian {CEO} problem
  under logarithmic loss,'' in \emph{IEEE IT Workshop}, 2018, pp. 1--5.

\bibitem{OptVGCEO}
J.-J. Xiao and Z.-Q. Luo, ``Optimal rate allocation for the vector gaussian
  {CEO} problem,'' in \emph{1st IEEE International Workshop on Computational
  Advances in Multi-Sensor Adaptive Processing}, 2005, pp. 56--59.

\bibitem{localiz}
A.~{Ababneh}, ``Low-complexity bit allocation for {RSS} target localization,''
  \emph{IEEE Sensors Journal}, vol.~19, no.~17, pp. 7733--7743, 2019.

\bibitem{fr1}
C.~{Huang}, R.~{Mo}, and C.~{Yuen}, ``Reconfigurable intelligent surface
  assisted multiuser {MISO} systems exploiting deep reinforcement learning,''
  \emph{IEEE Journal on Selected Areas in Comm.}, vol.~38, no.~8, 2020.

\bibitem{fr2}
C.~{Huang}, G.~C. {Alexandropoulos}, A.~{Zappone}, C.~{Yuen}, and M.~{Debbah},
  ``Deep learning for {UL/DL} channel calibration in generic massive {MIMO}
  systems,'' in \emph{IEEE International Conference on Communications (ICC)},
  2019, pp. 1--6.

\bibitem{Integer_opt}
B.~{Farber} and K.~{Zeger}, ``Quantization of multiple sources using integer
  bit allocation,'' in \emph{Data Compression Conference}, 2005, pp. 368--377.

\bibitem{kNN_dep}
S.~Gao, G.~V. Steeg, and A.~Galstyan, ``{Efficient Estimation of Mutual
  Information for Strongly Dependent Variables},'' in \emph{Proceedings of the
  18th International Conference on Artificial Intelligence and Statistics},
  vol.~38.\hskip 1em plus 0.5em minus 0.4em\relax San Diego, California, USA:
  PMLR, May 2015, pp. 277--286.

\bibitem{LQR}
\BIBentryALTinterwordspacing
W.~S. Levine, \emph{The Control Handbook}, ser. Electrical Engineering
  Handbook.\hskip 1em plus 0.5em minus 0.4em\relax Taylor \& Francis, 1996.
  [Online]. Available: \url{https://books.google.de/books?id=2WQP5JGaJOgC}
\BIBentrySTDinterwordspacing

\bibitem{moonturl}
\BIBentryALTinterwordspacing
``Classifier comparison with scikit-learn.'' [Online]. Available:
  \url{https://scikit-learn.org/stable/auto_examples/classification/plot_classifier_comparison.html}
\BIBentrySTDinterwordspacing

\bibitem{scikitlearn}
F.~Pedregosa, G.~Varoquaux, A.~Gramfort, V.~Michel, B.~Thirion, O.~Grisel,
  M.~Blondel, P.~Prettenhofer, R.~Weiss, V.~Dubourg, J.~Vanderplas, A.~Passos,
  D.~Cournapeau, M.~Brucher, M.~Perrot, and E.~Duchesnay, ``Scikit-learn:
  Machine learning in {P}ython,'' \emph{Journal of Machine Learning Research},
  vol.~12, pp. 2825--2830, 2011.

\bibitem{githubindoorenv}
\BIBentryALTinterwordspacing
M.~I. AlHajri, N.~T. Ali, and R.~M. Shubair, ``2.4 {GHZ} indoor channel
  measurements data set.'' [Online]. Available:
  \url{https://archive.ics.uci.edu/ml/datasets/2.4+GHZ+Indoor+Channel+Measurements}
\BIBentrySTDinterwordspacing

\bibitem{7696430}
M.~I. {AlHajri}, N.~{Alsindi}, N.~T. {Ali}, and R.~M. {Shubair},
  ``Classification of indoor environments based on spatial correlation of rf
  channel fingerprints,'' in \emph{2016 IEEE International Symposium on
  Antennas and Propagation (APSURSI)}, 2016, pp. 1447--1448.

\bibitem{indoorenvpaper}
M.~{I. AlHajri}, N.~T. Ali, and R.~M. Shubair, ``Classification of indoor
  environments for iot applications: A machine learning approach,'' \emph{IEEE
  Antennas and Wireless Propagation Letters}, 2018.

\end{thebibliography}
	
	
	%\vfill
	
	
	\section*{Impact of Bit Allocation Strategies on Machine Learning Performance in Rate Limited Systems, \underline{Extension}}
	
	In the following, we provide the numerical results for other scenarios covering different MLUs, regression and classification, and real- and complex-valued attributes. All the considered simulations show significant gains when using the proposed KLD method, demonstrating its power and  benefits when used in rate limited systems. This is also theoretically expected because the conventional methods like MSE do not take the final MLU decision into consideration. The aforementioned problems are listed below and their description and simulation results are provided afterwards. 
	
	\begin{enumerate}[label=(\roman*)]
		\item Moon data set
		\item Inverted pendulum with different setup
		\item 2.4 GHz indoor environment classification with vector quantization 
	\end{enumerate}

	\begin{enumerate}[label=(\roman*)]
		\item \textbf{Moon Data Set}
	\end{enumerate} 
	
\begin{figure}[t!] 
		\centering
		\includegraphics[width=.95\columnwidth]{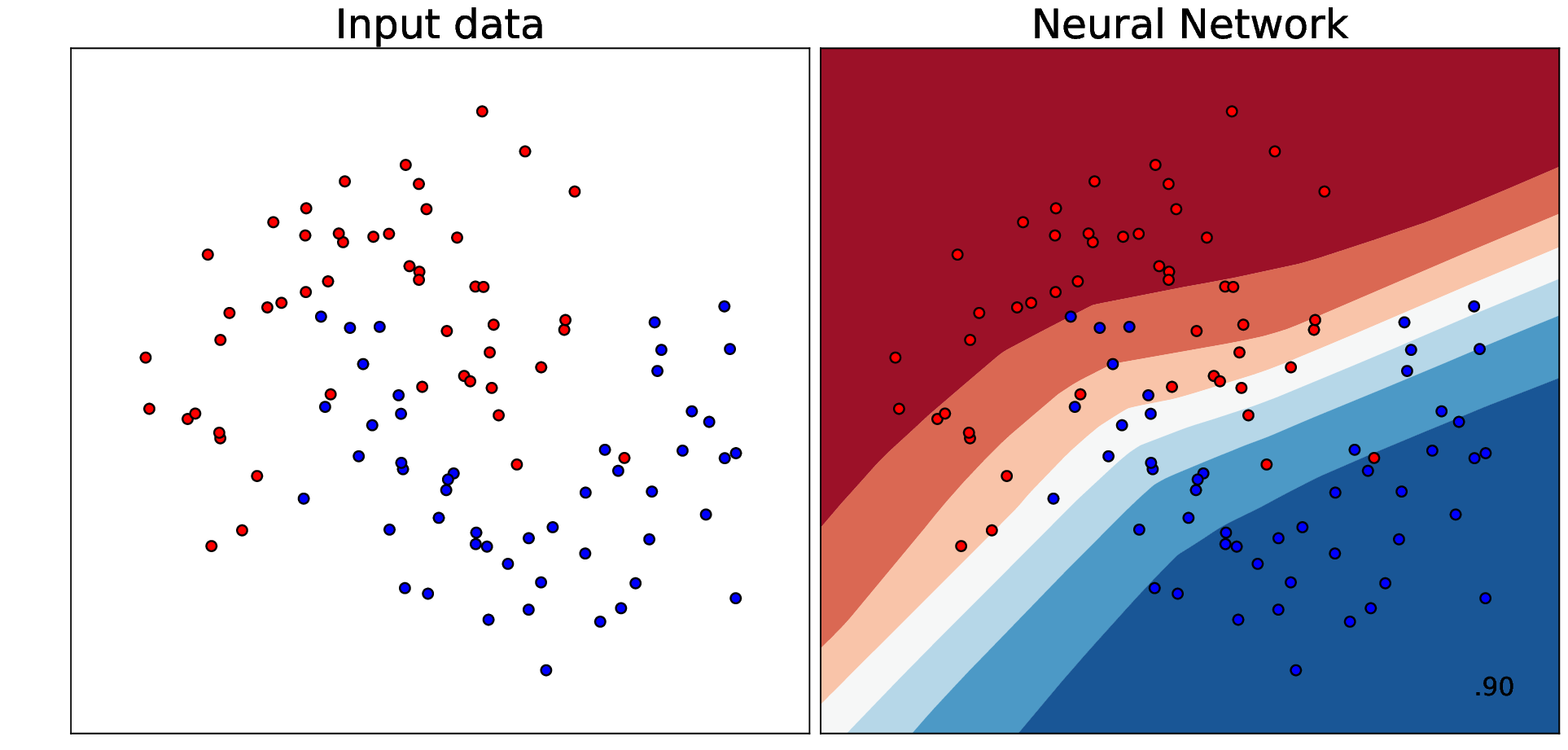}
		\vspace{-.2cm}
		\caption{Moon data set and NN Classifier with non-quantized data.} \label{moon}
	\end{figure}

	The moon data set is presented in scikit-learn to perform classification tasks (Figure~\ref{moon}). The data set and more details are available in~\cite{moonturl,scikitlearn}. Assuming $2 \leq R_i \leq 7$ to determine the feasible set $\mathcal{R}$, we get the results shown in Table~\ref{moont}.

	\begin{enumerate}[label=(\roman*)]
		\setcounter{enumi}{1}
		\item \textbf{Inverted Pendulum with Different Setup}
	\end{enumerate}
	
	In the manuscript, it is assumed that bar mass and length do not change frequently and thus, quantized with high accuracy (10 bits). Here, we assume the rest of RVs are quantized with high accuracy and the bit allocation is performed on bar mass and length: $m, l$. 
	
	$\mathcal{R}$ is defined for $1 \leq R_i \leq 7$ and $R_{\mathrm{sum}} =5$. The simulation results are shown in Table~\ref{mlt}. As it can be seen, the KLD approach picks a bit allocation which results in \textbf{achieving zero steady state errors}. For the same case, MSE picks a bit allocation to decrease the quantization noise on $m$ which has a larger interval, however the controller sensitivity to changes in $l$ is higher. Hence, the MSE selection results in a degradation of $\mathbf{16.2 \%}$ in performance. Equal sharing allocates 2 bits instead of just 1 bit for $l$ and thus, the performance loss becomes $\mathbf{1.6\%}$.

	For $R_\mathrm{sum} > 5$, $P_e = 0$ for both KLD and MSE methods. As mentioned in the manuscript, the method shows high gains for  systems with limited resources.
	
	\begin{table}[t!]
		\vspace{.9cm}
		\centering
		\begin{tabular}{ | c | c | c |} 
			\hline 
			& Selected bit allocation & Classification Accuracy (\%) \\ \hline
			The proposed KLD & 4, 3 bits & 92.5 \\ \hline
			MSE (benchmark) & 7, 7 bits & 90  \\  \hline
		\end{tabular} 
		\vspace{.1cm}
		\caption{Moon data set results. \textbf{\color{RoyalBlue} \underline{Brief conclusion:} The proposed KLD selects a bit allocation that results in both  $\mathbf{2.5\%}$ gain in classification accuracy and  $\mathbf{50 \%}$ gain in number of used bits comparing with the bit allocation selected by MSE.}} \label{moont}
	\end{table}

	\begin{table}[t!]
		\centering
		\begin{tabular}{ | c | c | c |} 
			\hline 
			& Selected bit allocation & \makecell{Steady state error\\ probability (\%)} \\ \hline
			The proposed KLD & 2, 3 bits & 0 \\ \hline
			MSE (benchmark) & 4, 1 bits & 16.2  \\  \hline
			Equal Sharing (benchmark) & 2, 2 bits & 1.6 \\ \hline
		\end{tabular}
		\vspace{.1cm}
		\caption{Results for simulations with different inverted pendulum setup, quantizing bar mass and length. \textbf{\color{RoyalBlue} \underline{Brief conclusion:} For $R_\mathrm{sum} \leq 5$, the proposed KLD achieves the best performance of $P_e \mathbf{= 0}$, indicating a gain of $\mathbf{\approx 16\%}$ and $\mathbf{1. 6\%}$ comparing with MSE and equal sharing methods.}} \label{mlt}
	\end{table}

	\vspace{.5cm}
	\begin{enumerate}[label=(\roman*)]
		\setcounter{enumi}{3}
		\item \textbf{2.4 GHz Indoor Environment Classification with Vector Quantization}
	\end{enumerate}
	
	The 2.4 GHz indoor environment classification data set is available in~\cite{githubindoorenv, 7696430}. Here, we assume that channel transfer function (CTF) and frequency coherence function (FCF) attributes are transmitted to the MLU from two terminals. Each of the CTF and FCF vectors have 601 complex-valued thus, 1202 real-valued attributes. For more details, see~\cite{indoorenvpaper}.
	
	In this part, we apply kmeans as quantization on CTF and FCF vectors . The simulation results for $4 \leq R_i \leq 8$ are shown in Table~\ref{indoorVQ}. It can again be observed that the proposed KLD method provides the \textbf{best classification performance, e.g., 10\% and 7\% gain} compared to MSE and equal sharing for $R_\mathrm{sum} = 10$, respectively.

	\begin{table}[t!]
		\centering
		\begin{tabular}{ | c | c | c |c|} 
			\hline 
			$R_\mathrm{sum}$ & The proposed KLD & \makecell{MSE \\ (benchmark)} & \makecell{Equal sharing \\(benchmark)} \\ \hline
			10 bits &  69 \% & 59 \% & 63 \%\\ \hline
			14 bits & 82.8 \% & 77.7\% & 78 \%  \\  \hline
		\end{tabular} 
		\vspace{.1cm}
		\caption{Classification accuracy (\%) for 2.4 GHz indoor environment classification with NN and vector quantization. \textbf{\color{RoyalBlue} \underline{Brief conclusion:} The proposed KLD delivers the highest classification accuracy for different constraints on total number of used bits, showing a gain of at least $\mathbf{\approx 5\%}$ to $\mathbf{10\%}$ comparing with other methods.}} \label{indoorVQ}
	\end{table}

\end{document}